# Harnessing the Benefits of Open Electronics in Science

**Michael Oellermann[1,2]\*, Jolle W. Jolles[3], Diego Ortiz[4], Rui Seabra[5], Tobias Wenzel[6], Hannah Wilson[7], Richelle Tanner[8]**

[1] Technical University of Munich, TUM School of Life Sciences, Aquatic Systems Biology Unit, Weihenstephan, Germany; [2] University of Tasmania, Institute for Marine and Antarctic Studies, Fisheries and Aquaculture Centre, Hobart, Australia; [3] Centre for Research on Ecology and Forestry Applications (CREAF), Campus UAB, Bellaterra Barcelona, Spain; [4] INTA, Instituto Nacional de Tecnología Agropecuaria, Estación Experimental Manfredi, Manfredi, Argentina; [5] CIBIO-InBIO, Centro de Investigação em Biodiversidade e Recursos Genéticos, Universidade do Porto, Campus Agrário de Vairão, Portugal; [6] Pontificia Universidad Católica de Chile, Institute for Biological and Medical Engineering, Schools of Engineering (IIBM), Medicine and Biological Sciences, Santiago, Chile; [7] Utah State University, College of Science, Biology Department, Logan, UT, USA; [8] Chapman University, Schmidt College of Science and Technology & Wilkinson College of Arts, Humanities, and Social Sciences, Environmental Science and Policy Program, Orange, CA, USA

\* Corresponding author: M. Oellermann (michael.oellermann@utas.edu.au)

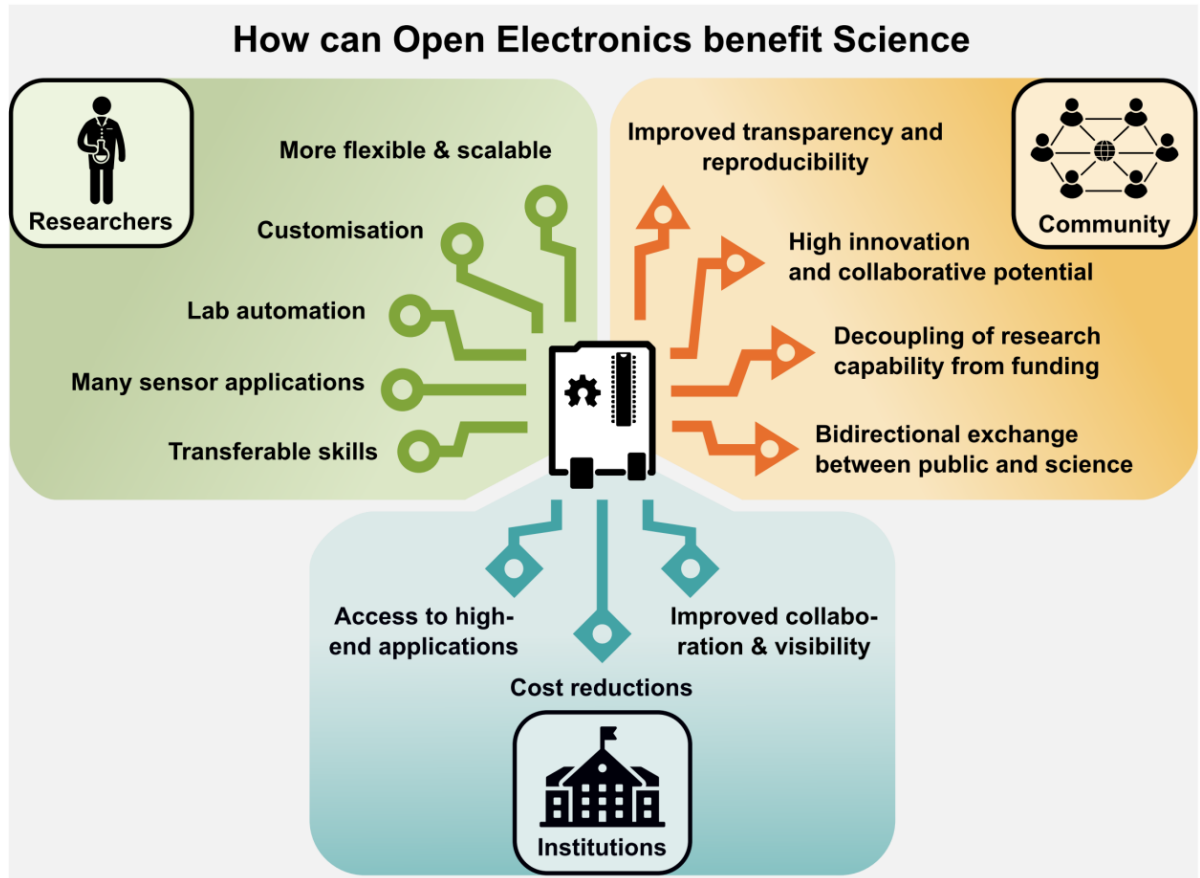

**Abstract**

Freely and openly shared low-cost electronic applications, known as open electronics, have sparked a new open-source movement, with much un-tapped potential to advance scientific research. Initially designed to appeal to electronic hobbyists, open electronics have formed a global community of "makers" and inventors and are increasingly used in science and industry. Here, we review the current benefits of open electronics for scientific research and guide academics to enter this emerging field. We discuss how electronic applications, from the experimental to the theoretical sciences, can help (I) individual researchers by increasing the customization, efficiency, and scalability of experiments, while improving data quantity and quality; (II) scientific institutions by improving access and maintenance of high-end technologies, visibility and interdisciplinary collaboration potential; and (III) the scientific community by improving transparency and reproducibility, helping decouple research capacity from funding, increasing innovation, and improving collaboration potential among researchers and the public. Open electronics are powerful tools to increase creativity, democratization, and reproducibility of research and thus offer practical solutions to overcome significant barriers in science.

## 1. Introduction

The revolutionary open science movement has helped to foster transparency, collaborative access, and sharing of scientific knowledge (Vicente-Saez and Martinez-Fuentes, 2018). Open science started with open-access publications and has now expanded to liberate access to data, program code, and even lab notebooks (Boulton et al. 2012; McCray et al. 2018; Vicente-Saez & Martinez-Fuentes 2018). However, so far one domain, which is at the very core of scientific data production, has been missing in the open science movement: hardware, electronics, and instruments (Harnett 2011; Pearce 2012; Maia Chagas 2018). Cutting-edge instruments enable high-profile research, yet high costs limit their access only to well-funded labs. The majority of researchers globally do not have access to the funding required to buy state-of-the-art instruments, limiting both reproducibility and innovation potential (van Helden 2012). Free and Open-Source Hardware (Pearce 2013) has the potential to close this divide: it facilitates sharing of free design blueprints to re-build, modify, or advance instruments and foster collaboration with other scientists and a worldwide community of "makers", civic scientists, and hobbyist inventors (Pearce 2012; Maia Chagas 2018).

Open electronics are a major component of the open hardware domain, which provides open-source scientific hardware solutions (Pearce 2012; Bonvoisin et al. 2020). Open instrumentation solutions are often built on electronic hardware components (some are open source themselves) whose main purpose is to allow non-experts to easily create electronic applications. This includes single-board microcomputers (SBCs) and micro-controllers, and a plethora of inter-compatible hardware modules, sensors, actuators, and displays (Figure 3, Table 1) that can be easily interfaced with each other, many with little prior experience required. In combination with the modular nature of many open electronics platforms such as that of the popular Raspberry Pi and Arduino, users do not need to invent applications from scratch and can gradually grow skills and application complexity. Despite the need for some basic programming and electronics skills, open electronics projects are now even accessible to pre-school children, supported by a vast number of open online tutorials and databases (e.g., instructables.com; hackster.io). With millions of hobbyist makers and DIYers around the globe, and more than 37 million Raspberry Pi microcomputers sold till Jan 2020 alone (Raspberry Pi Foundation 2020; Upton 2020), the popularity of open electronics has continued to rise and is beginning to establish in diverse scientific domains (Figure 1A, B; Jolles 2021a).

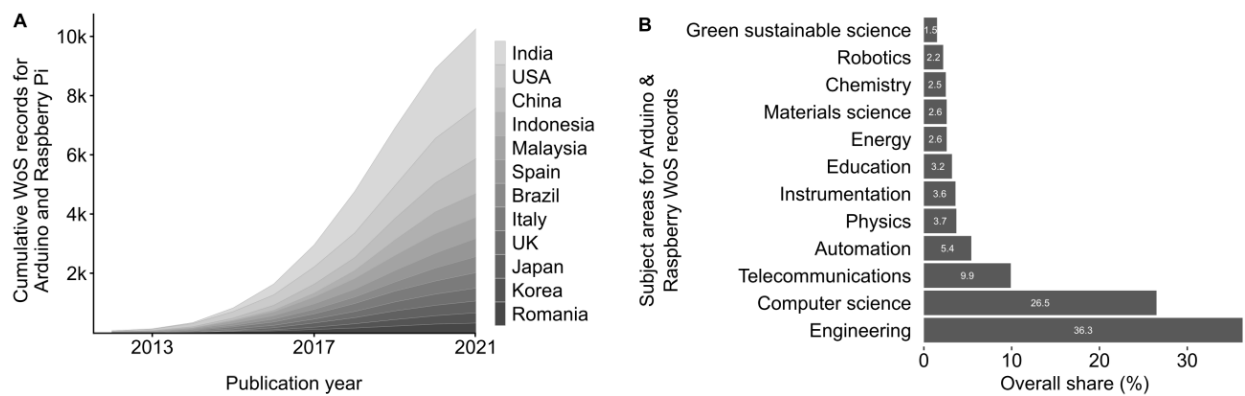

**Figure 1. (A)** Cumulative growth of Web of Science records grouped by the top 12 countries and **(B)** dominant subject areas for the search terms "Arduino" and "Raspberry Pi", for author and co-author origins from 2010 – 2020. Articles and proceeding papers were pooled. For detailed analysis, data and country distribution for proceedings articles only see Supplementary Information File S1.

Despite its increasing uptake in science, open electronics applications are far from being widespread. Poor awareness, rare documentation, and insufficient electronic literacy outside the engineering and computer sciences has contributed to its fragmented and uneven use across scientific subjects (Figure 1B). However, to develop standards (Bonvoisin *et al.* 2020), best practices and foster innovation, open electronics will need to become common tools in experimental research. In comparison, open-source software projects such as the R statistical language have shown to be truly innovative and adapted rapidly to new demands and research trends, via user-driven innovation networks (Von Hippel 2005; Von Hippel 2007), lifting it to one of the most popular data tools in science (Muenchen 2012; Lai *et al.* 2019). Open electronics have comparable potential for science but face significant barriers, such as lacking awareness of their multiple benefits and a widespread open-sharing culture, to foster iterative and collective advancements of experimental applications. Such barriers can be broken down, by making information increasingly available, such as detailed construction blueprints, troubleshooting guides and safety standards (Murillo & Wenzel 2017) and by presenting a clear case for how open electronics can benefit researchers, institutions and the scientific community alike. This will help to accelerate hardware innovation, democratize hardware access, lower research costs, and enable highly customizable solutions for experimental science (Powell 2012; Pearce 2015; Pearce 2016).

Here we outline the broad benefits that open electronics can have for researchers, institutions, and the scientific community at large. We then discuss current barriers and provide a "Beginners Toolbox" to help researchers get started and conclude with an outlook discussing their potential impacts on science and academia and the actions required to foster a broad uptake. Overall, we aim to raise positive awareness about the multiple benefits of open electronics and thereby promote innovation, reproducibility, and democratization of science.

## 2. Application potential for open electronics in science

Open electronics offer a hugely versatile spectrum of applications to a wide range of potential users in science, education, industry, and the general public. Although initially used only by the most electronics-savvy hobbyists and Do-it-Yourself creators, open electronics are increasingly taken up by broader public audiences that span all age groups, further fueled by the rise of the Internet of Things (Ibrahim *et al.* 2015). Automation of scheduled tasks such as watering plants in the garden (Divani, Patil & Punjabi 2016) or controlling household devices (i.e., smart homes) are very popular and easy to set up (Hasan *et al.* 2018). This extends to various measurements and surveillance applications (e.g., weather stations or birdhouses) and even for developing smart cities (Costa & Duran-Faundez 2018). Some of the driving forces behind the rise of open electronics was to bring computing and electronics to the broader public and make it accessible to anyone. This has started to cross over to STEM education where hands-on experience on building devices can be used for introducing students with electronics and programming basics as well as solving practical problems and practice the scientific method (Jolles 2021a). The increasing interest in open electronics as teaching tools is supported by an extensive pool of learning resources for teaching or self-learning (see Table 2 in beginners toolbox). This is not only useful for individuals and smaller companies who cannot afford professional development of electronic components for their prototypes, but also for scientists who want to test new ideas or customize experiments on a small budget.

So far, there has been only a marginal uptake of open electronics in science, with predominant use in the engineering and computer sciences (Figure 1B), despite much potential for applications to span the full spectrum of scientific disciplines. Examples from the biological sciences include the behavioral video-monitoring of woodpeckers (Prinz *et al.* 2016), honeybees (Ai *et al.* 2017) or zebrafish (Maia Chagas *et al.* 2017), automated bird feeders (Philson *et al.* 2018), RFID based automated weight measurements of mice (Noorshams, Boyd & Murphy 2017), underwater video surveillance (Mouy *et al.* 2020), and the remote measurement of body temperature and respiratory rate of mice (Kallmyer *et al.* 2019). Agricultural sciences have used open electronics to monitor e.g. plant disease (Gonzalez-Huitron *et al.* 2021), nutritional status (Brambilla *et al.* 2021) or environmental variables such as relative humidity, temperature, light or dissolved oxygen in plant factories (Montoya *et al.* 2020). And social scientist have for example used open electronics to study sentiments in social media (Alzahrani & Ieee 2018), perceptual illusions (Ferracci & Brancucci 2019) or auditory distractions on cyclists (Scanlon *et al.* 2020). Examples extend further to other disciplines including chemistry (Urban 2015), health sciences (Pitarma, Marques & Ferreira 2017) and astronomy (Ferkinhoff 2014), highlighting the remarkable flexibility and broad application potential of open electronics in science. In addition, open electronics are formidable tools for citizen science and scientific outreach activities such as school student-operated ocean observers (Beckler *et al.* 2018), urban air pollution monitors (Jiang *et al.* 2016) or sonic kayaks to monitor underwater soundscapes (Griffiths *et al.* 2017).

## 3. Benefits of open electronics for scientific research

In addition to their diverse application potential, open electronics can provide a broad range of significant benefits at the different levels of academia and resolve important practical, financial, and structural issues.

### 3.1 Benefits to individual researchers

*Wide applicability, from simple to complex*

Unlike most scientific instruments, open electronics are highly flexible and adaptable, and can be implemented in a broad range of applications, from basic to highly complex, including closed-loop operant chambers (O'Leary et al., 2018) and real-time virtual reality systems (Tadres and Louis, 2020). Users can start simple and expand their devices with increasing programming and electronics skills, such as starting with only logging lab temperature, then displaying it live on an LCD screen, controlling heaters to regulate temperature, to finally, a complete stand-alone system with multiple sensors, warning messages, and interactive graphical user interfaces. Users can also easily repurpose open electronics by reusing components from previous setups for new or more complex builds.

### Broad sensor and actuator application potential

A major strength of open electronics is the wide range of sensors and actuators available that can be controlled with the accuracy of reference equipment (Table 1; Setyowati, Muninggar & Shanti 2017). Open electronics can also be used in applications with a very small footprint (e.g. Palossi, Conti & Benini 2019) both in the lab and under harsh conditions in the field (e.g. Beddows & Mallon 2018). Micro-controllers and single-board computers also enable multiple sensors and actuators to be connected simultaneously, providing much greater sensing and reactive capacity than most commercial devices while significantly reducing equipment needs, costs and power consumption.

### Lab automation

Repetitive tasks, such as control and recording of experimental parameters, mixing reagents, animal feeding, and monitoring of experimental trials, are amongst the most time-consuming factors in research labs. Open electronics can benefit researchers by automating such tasks, including by using robotics for pipetting (Steffens *et al.* 2017; Florian *et al.* 2020), RFID-based animal feeding stations (Bridge *et al.* 2019), or smart IoT monitoring systems generating high-density data streams to the cloud (Sethi *et al.* 2018; Arunachalam & Andreasson 2021). Task automation also helps reduce human error and experimental variability (Eggert *et al.* 2020) and increases resilience to unforeseen circumstances.

Table 1. Overview of the huge range of available open electronics sensors and actuators compatible.

| Sensors |
| --- |
| *Environment* (temperature, humidity, barometric pressure, soil moisture, particulate matter, light intensity, smoke, dust, radiation) |
| *Movement* (distance, acceleration, seismic, GPS, break-beam, motion) |
| *Gas* (CO, $CO_2$, alcohol, $H_2$, TVOCs, ozone, $H_2S$, $CH_4$, NO) |
| *Biometrics* (heart rate, muscle activity, fingerprints, weight/load, force) |
| *Water* (chlorine, pH, depth and pressure, liquid level, flow, turbidity) |
| *Imaging* (spectroscopy, visible and IR range cameras, thermal imaging, gestures) |
| *Other* (magnetism, capacitive touch, current, voltage, sound, RFID) |

| Actuators |
| --- |
| *Switches* (mechanical, electrical, magnetic, DC and AC relays) |
| *Movement* (servos, stepper motor, gear motor, vacuum pumps, valves) |
| *Light* (LEDs, infrared, UV, laser) |
| *Other* (vibration, sound, ultrasound, Peltier heating/cooling) |

## Scalability and high throughput

Open electronics provide researchers with the opportunity to easily scale and replicate setups to suit singular or high-throughput applications. Their low cost and off-the-shelf availability enables quick and low-risk prototyping up until a well-functioning setup that can be copied to create whole arrays of identical devices, such as to GPS-track tens of animals (Foley & Sillero-Zubiri 2020), test the behavior of hundreds of individual flies (Geissmann *et al.* 2017), to observe the growth of thousands of plants (Tausen *et al.* 2020), and to parallelize automated processing of sample microvolumes (microfluidics) for microbiology and single cell RNA sequencing (Stephenson *et al.* 2018; Wong *et al.* 2018). Such scalability is particularly valuable when funding is limited, enabling researchers to begin with simpler setups, rather than facing high upfront costs for commercial systems.

## Customization

Most instruments, such as HD cameras, plate readers, microscopes, and PCR machines, are closed entities, constrained to the functions set by the manufacturer and operating software, and can thus become redundant if research needs change. The poor ability to modify or expand functionalities also confines the scope and implementation of new research ideas. Open electronics can solve this as researchers can not only develop or retrofit existing open electronics setups and devices, exchange, or program new operations, but also link and expand the features of existing laboratory instruments. For example, microcontrollers and SBCs can interface with commercial instruments via serial ports and hardware communication protocols, to query information or execute functions, while adding new functionalities using sensors and actuators (e.g. Rodríguez-Gómez *et al.* 2019; Arce & Stevens 2020; Virag *et al.* 2021). With ever newer generations of boards focused on facilitating Internet-of-Things applications (e.g., Adafruit Feather HUZZAH, Figure 3), even simple weight scales can integrate into a smart instrument network, channeling and summarizing data streams in cloud-based dashboards (Poongothai, Subramanian & Rajeswari 2018; Arunachalam & Andreasson 2021).

## Flexible data access and programming capabilities

Open electronics are highly flexible in terms of data acquisition, formats, storage, and accessibility. Numerous libraries in a broad range of programming languages make it possible to read sensor data in a few lines of code. Library-rich programming languages such as Python further facilitate endless possibilities to work with custom electronics and devices, including automatic data processing actions such as folder monitoring, file conversion and automatic creation of data backups. Data can also be accessed remotely, including to a local network and the internet, and from remote field locations via mobile network adaptors (e.g. Sethi *et al.* 2018). This in turn enables the continuous real-time remote monitoring of data, such as of lab conditions, animal activity, plant growth, and environmental variables in the field (Siregar *et al.* 2017; Trasviña-Moreno *et al.* 2017; Jolles 2020). Improved computing power of SBCs has made it increasingly possible to process data onboard, enabling only the temporal transmission of flagged or summarized data for researchers (Allan *et al.* 2018). Data can also be professionally visualized via user interfaces or online dashboards supported by numerous graphical libraries, of which most are open-source itself (e.g., Tkinter, PyQT, WxPython, dash, plotly (Boudoire *et al.* 2020; Lewinski *et al.* 2020).

## Easy to service and troubleshoot

Most components of open electronics can be easily serviced and replaced by the users themselves, with most parts likely to be available at online retailers and electronic hardware stores. Also, required tools, such as soldering equipment and a multi-meter, tend to be highly affordable. In contrast, when issues occur with commercial (scientific) instruments, custom repairs, even when feasible, are not recommended as they break product warranty. Researchers therefore rely on manufacturers for repairs, which can be time-consuming and potentially risky as support may cease when products become outdated.

*Extensive learning resources and community support*

Extensive learning resources, including a large range of books and free tutorial websites (see Table 2), and an increasing number of open online courses (e.g. Coursera, Udemy) offer many ways to learn about open electronics and how to build custom applications. Academic papers now often come with supplementary guides and accompanying websites about methodologies (e.g. Geissmann *et al.* 2017; Minervini *et al.* 2017; Maia Chagas 2021), and a number of specialized journals exist to help researchers build and publish their own devices (e.g. Journal of Open Hardware, HardwareX). It is also easier to troubleshoot problems, as most open electronics applications are built on similar and wide-spread building blocks (i.e. Arduino platform) that share a common programming language, and large online communities exist that can be consulted to help solve specific problems (e.g. stackoverflow.com and raspberrypi.org/forums, which has 300k+ members).

*Transferable skills*

Besides providing practical benefits, learning to work with open electronics and creating custom devices and applications also provides researchers with transferrable skills, including knowledge of programming and electronics, and creative thinking, which is paramount to scientific progress. It also enables researchers to be more independent from funding constraints and access to commercial vendors' support.

### 3.2 Benefits to departments and institutions

*Access to high-end applications*

Open electronics as an approach to provide cutting-edge scientific instrumentation has matured quite considerably over the last few years. There are numerous examples of open electronics instruments with uncompromising quality being used for high-end scientific research, such as magnetic resonance tomography (Moritz *et al.* 2019), automated microbiological incubators (Wong *et al.* 2018), high-throughput tracking and optogenetic stimulations (Tadres & Louis 2020), and microfluidic single cell sequencing preparation (Stephenson *et al.* 2018). It is becoming increasingly advantageous for academic institutions to adopt open electronics solutions as a leaner way to perform workflows in-house through a modular, gradual investment, overcoming the need for researchers to depend on large grants. Institutions can facilitate this by providing dedicated open-electronics workspaces, where researchers cannot only implement their own ideas, but also form institutional networks to share knowledge, ideas and instruments across departments and stimulate interdisciplinary innovation.

*Equipment maintenance and extension*

When encouraged as an institutional-wide policy, the cost effectiveness of open electronics can be extended throughout the lifetime of equipment. Maintenance is enabled through open hardware documentation and the knowledge pool that naturally emerges within the staff during assembly and operation of open electronics instruments. Both ensure that the majority of maintenance and repair operations can also be performed in-house quickly with minimal fabrication expenses beyond parts. Additionally, this approach is more environmentally sustainable than proprietary solutions. Such control over the fate of critical scientific equipment is crucial for all research institutions, but especially so for institutes in countries where local technical support from commercial vendors may be lacking or prohibitively expensive.

While small custom setups are most widely represented among open electronics projects, the advantages are by no means limited to these. For example, in order to grow and maintain their large infrastructure sustainably, the European Center for Nuclear Research (CERN) builds the electronic components of the particle accelerator with open source hardware (van der Bij *et al.* 2012). Following this uncommon path, they have been simultaneously innovating in commercial sub-contract formats, electronics CAD software KiCAD, and the CERN Open Hardware License (Svorc & Katz 2019). One resource example which was developed in this context but is now used across academia and industry, is the White Rabbit, the current gold standard to achieve ultra-fast data transfer synchronization in Ethernet networks (Moreira *et al.* 2009).

### Improved collaboration and visibility

Open electronics can link virtually all fields of research. At the institutional level, collaborations can be facilitated around the development and implementation of open electronics solutions for frontier research applications. This could be fostered by intra- and interinstitutional think-tanks, workspaces, and shared educational programs, and complemented by technical support where researchers lack the required electronics or programming expertise. This collaborative approach also enhances publications, where useful tools are published in addition to research data and is likely to affect citation rates positively in similar ways as shown for open data (Colavizza *et al.* 2020). A clear commitment to technologies that democratize science will also help institutions to enhance collaborations between industrialized and emerging nations and attract researchers that can easily cross-transfer open electronics technologies. Potentially, this will not only improve institutes' international visibility and reputation but may also help in acquiring public funding.

### 3.3 Benefits to the scientific community and funders

### Improved transparency and reproducibility

Transparency and reproducibility are hallmarks of the scientific method, but prohibitively high costs and lack of documentation of procedures and tools in published methods commonly prevent effective replication. Open electronics offer an opportunity to counter this issue. Published works based on open electronics become technically and financially easier to reproduce through decreased reliance on proprietary solutions. At the same time, as the development of open electronics instruments becomes increasingly publishable, researchers are incentivized to transparently communicate the details of the solutions employed.

### Decoupling of research capability from funding

The lean nature of open electronics enables specialized research and is much more conducive to experimentation and exploration than most commercial solutions. For example, the use of electronics in biological research in harsh ecosystems, such as wave-swept rocky shores or remote deserts, is difficult, and equipment may easily become damaged or lost. In this context, researchers either secure more funds to cover the losses of expensive material or down-scale the research line. Alternatively, examples show that open electronics can be efficiently harnessed to develop cheaper and fully fit-for-purpose equipment (Burnett *et al.* 2013; Gandra, Seabra & Lima 2015), while minimizing the cost incurred when losses occur. These and equivalent solutions alleviate the entry cost of many research topics and contribute to a greater decoupling of research capability from funding, ultimately facilitating the exploration of novel research lines and supporting investigations of early career researchers and scientists worldwide that have reduced access to infrastructure and funding.

### High innovation and collaboration potential

It is a common prejudice that open-source development conflicts with commercialization and industry collaboration. In reality, just like successful open-source software companies, open electronics is an excellent basis for commercial knowledge transfer. Research-driven technological innovation involves developers (typically engineering-oriented teams but increasingly also open electronics makers) and end-users (typically non-engineering-oriented researchers). More often than not, the greatest obstacle to the innovation process is ineffective communication between both groups and user centered design. With an unparalleled wealth of learning resources and inexpensive entry-level equipment, open electronics represents the ideal method for scientists to become fluent in basic electronics and programming and to foster communication and collaboration between developers and scientific users. Increased technological literacy of users additionally ensures that end-users have a better grasp of current technological boundaries, permitting the establishment of goals that are simultaneously realistic and ambitious. At the same time, this lean development approach speeds up development cycles that often result in fully functional solutions, and in many cases is further enhanced by free user contributions. Those and further advantages (e.g. fast-adaptation, easy user engagement and advertisement) can outweigh the disadvantages of such open source business models (e.g. reduced profit timeframe after innovation cycles are stopped, less acceptance of excessive price margins) and provide rewarding opportunities for commercial developers and scientific users alike (Pearce 2017). At the user´s level, increasing adoption of open electronics means skill development relevant for science and industry employment even in fields where such skills have not been traditionally taken into consideration, such as the biological sciences (Jolles 2021a).

*Bidirectional knowledge transfer between public and science*

While an increasing number of scientists has been inspired by the large pool of freely-shared open electronics solutions (e.g. by home applications such as surveillance and home automation) to integrate those solutions into scientific experiments (Jolles 2021a), it also offers great opportunity to facilitate bidirectional collaborations with the public and science. Funders and society increasingly expect scientists to engage more actively with the public to improve the uptake and application of scientific knowledge (Hunter 2016). At the same time there is an increasing demand by the public to actively engage in the scientific process, to an extend that citizens partner even co-author with professional scientists (Breen *et al.* 2015; Mazumdar, Wrigley & Ciravegna 2017). However, access to scientific instruments has partly hampered this process as well as bottom-up approaches where citizens themselves develop scientific questions (Mazumdar, Wrigley & Ciravegna 2017; Ostermann-Miyashita, Pernat & König 2021). Open electronics can overcome this barrier by providing cost-effective and interactive tools that can be easily rebuild by non-experts, while providing high-quality scientific data (Weeser *et al.* 2018). The lean and modular design that is inherent to open electronics solutions further enhances a smooth exchange of knowledge and technical solutions between professional and civic scientists. Thus, open electronics are well suited to make science broadly reproducible and more accessible for new collaboration opportunities.

## 4. Barriers

It is clear there is remarkable potential for open electronics in science and academia. However, to reach this potential and reap the benefits at a broad scale, significant educational, collaborative, and technical barriers need to be overcome. Most researchers still lack basic awareness of the application potential, and the diversity of open electronics techniques and equipment. This is also clear from a recent review of the uptake of Raspberry Pi's in the biological sciences (Jolles 2021a), which identified a high number of different applications but with still limited uptake of such applications by different research groups. A major reason for this is the limited documentation of open electronics setups in scientific publications, which confines its visibility and the formation of any substantial academic Maker community (Glenn & Alfredo 2010; Harnett 2011). Instead, many open electronics techniques are spread among collaborators in an informal fashion. Initiatives exist that aim to increase the visibility of open hardware solutions, such as the Open Neuroscience network (Maia Chagas 2021), or new journals documenting open hardware designs in a systematic fashion, such as the Journal of Open Hardware and HardwareX. Nevertheless, poor awareness remains to be a significant barrier and hampers the broader academic community to use and thereby reproduce, re-create, and increase the visibility of open electronics solutions. Another barrier is the fragmentation of the existing open electronics community of users in academia, within institutions and across countries and subject domains, hindering knowledge exchange (Figure 1A, B). Within institutions and departments, there is often little support infrastructure for educational resources and community-building, such as institutional user-run Maker workshops (Maia Chagas 2018). Across countries, use of open electronics in laboratories is concentrated in non-western nations (Figure 1A), which does not mirror international collaboration networks, dominated by the USA and Europe (Hennemann, Rybski & Liefner 2012). Thus, knowledge flow is limited at a global scale and often enclosed locally.

A further hindrance is the uneven use and recognition across scientific subjects and disciplines. Unsurprisingly, engineers publish most frequently with the explicit terms "Arduino" and "Raspberry Pi" (36.3% of publications in our Web of Science search), followed by Computer Science (26.5%), Telecommunications (9.9%) and Automation (5.4%, Figure 1B, Supplementary Information File S1). Interestingly, a more detailed full-text analysis for the same search terms across all PLOS journals, showed that biological sciences form by far the major user base for these popular open electronics devices (33%, $n$=85), compared to engineering (11%, $n$=31) or computer sciences (4%, $n$=11). Yet only 3.5% of articles in the biological sciences reported their use in the abstract, in contrast to 19.4% in the engineering- or 18.2% in the computer sciences (for details of full text semantic analysis see Supplementary Information File S2), indicating a discipline-dependent bias to report open electronics applications. This may be due to open electronics applications being more visible in areas where methodologies are in focus (e.g. engineering) rather than non-technical research questions (e.g. biology). Eventually, authors do not always mention clearly or at all if they applied open electronics in their research. With improved acknowledgment researchers will recognize its value at a broader scale and potentially generate more associated research, innovation, and public interest.

Without an institutional and global collaborative community and widespread awareness, and the confidence to use and highlight open electronics designs, a broader uptake as well as more sophisticated developments will remain limited. Overcoming these barriers begins with increased visibility of the tools themselves and a cultivation of community around their use, to build confidence and electronics literacy. Global networks such as the Gathering for Open Science Hardware (Murillo *et al.* 2018), and an increasing number of scientific societies hosting dedicated symposia (e.g. Annual Meeting of the Society for Experimental Biology), are an excellent start to share innovative open electronics solutions across disciplines and budgets.

## 5. Beginners Toolbox

The best way to get started with open electronics is to dive right in and start building simple systems and applications, as this will give first-hand experience in how open electronics work and encourages the creative thinking that may lead to innovative applications (Figure 2). Hobby electronics starter kits provide great value for money and come with a large range of sensors, actuators, LEDS, breadboard, cables, and resistors that can be used to start tinkering. These are widely available online and can be used with both microcontrollers and single board computers (SBCs), for which a large range of options exist (see Figure 3 for an overview of devices). For beginners especially, Arduino and Raspberry Pi are recommended as they are the most popular and have by far the most documentation and support available. Because of their low cost, one can easily buy one of each and start to learn about the pros and cons of both devices. As first learning resources, a wide range of tutorials are available online that are geared towards hobbyists and teach fundamental electronic skills such as wiring, powering, and soldering when needed (Table 2).

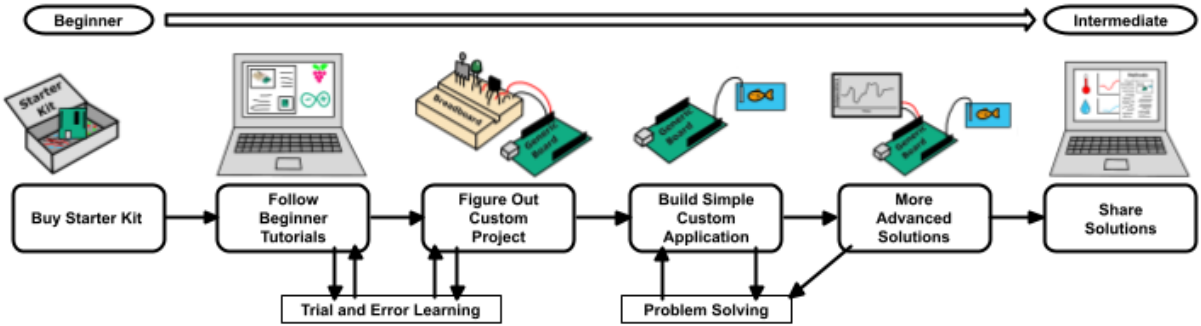

**Figure 2.** Diagrammatic representation of the potential steps for incorporating open electronics into ones' research. It is best to begin with a starter kit to explore its potential. Tutorials are useful for building initial skills, such as to set up a sensor to measure the temperature of an aquarium. Delving further into the many (online) resources available, basic systems can be expanded to perform more advanced tasks, such as plotting the temperature data in real time on a simple website and sending warning emails whenever values cross thresholds. The system can then be easily and affordably replicated and shared with the broader community.

| Board | Type | Price | Good for beginners | Performance | Flexibility | Resources & support | Machine Learning |
|---|---|---|---|---|---|---|---|
| Arduino Mega 2560 | MC | | | | | | |
| Arduino Uno R3 | MC | | | | | | |
| Asus Tinker Board | SBC | | | | | | |
| BeagleBone Black W | MC/SBC | | | | | | |
| Adafruit feather HUZZAH | MC | | | | | | |
| Nvidia Jetson Nano | SBC | | | | | | |
| Odroid C4 | SBC | | | | | | |
| PocketBeagle | SBC | | | | | | |
| Raspberry Pi 4B | SBC | | | | | | |
| Raspberry Pi Pico | MC | | | | | | |
| Raspberry Pi Zero W | SBC | | | | | | |
| Red Pitaya STEMlab | SBC | | | | | | |
| UDOO Bolt v8 | SBC | | | | | | |

**Figure 3.** Overview of some of the key micro-controllers and single-board computers on the market rated for their price, skill level, performance, flexibility, resources and support available, and possibility to run machine learning applications.

**Table 2**. Collection of online resources, for both beginners and advanced users, with hyperlinks. Many companies that sell components for open electronics provide thorough documentation and tutorials, including Arduino and Raspberry Pi. Furthermore, there are guides specifically developed for scientists wanting to work with the Raspberry Pi or Arduino (Jolles 2021b), and an increasing number of online courses are available on topics related to electronics and computing. The links below are arranged by relevance -- starting with beginner tutorials and ending with ways to share own applications.

| Resource | Link | Description |
|---|---|---|
| Arduino Website | arduino.cc | Many tutorials, forums, blog posts and products for sale |
| Raspberry Pi Website | projects.raspberrypi.org | Many tutorials, forums, blog posts and products for sale |
| Raspberry Pi Beginner's Guide | magpi.raspberrypi.org/books/beginners-guide-4th-ed | Free guidebook for getting started with Raspberry Pi |
| Adafruit Website | learn.adafruit.com | Thorough documentation and tutorials for Adafruit products |
| Sparkfun Website | learn.sparkfun.com | Thorough documentation and tutorials for Sparkfun products |
| PiHut Website | thepihut.com/blogs | Thorough documentation and tutorials for PiHut products |
| Raspberry Pi Guide | raspberrypi-guide.github.io | A collection of 30+ Raspberry Pi tutorials specifically written for scientists (Jolles, 2021) |
| Coursera | Coursera.org | Offers courses on topics related to electronics and computing |
| Udemy | Udemy.com | Offers courses on topics related to electronics and computing |
| TinkerCad | tinkercad.com | Lets you build virtual versions of circuits to test your wiring and code |
| Raspberry Pi Forums | raspberrypi.org/forums | 300k+ member forum to ask questions about Raspberry Pi |
| Open Hardware Science Forum | forum.openhardware.science | Forum to ask questions related to open hardware |
| Stack Overflow | stackoverflow.com | Forum to ask questions about hardware or coding |
| Open-Neuroscience | open-neuroscience.com | Database for scientific open-hardware designs |
| Github | Github pages | Site to create a free, version controlled online website with your documentation |

With some basic electronic and programming skills, open electronics can be easily integrated into most experimental designs, and researchers may benefit from the online resources and potential component lists provided in academic papers and online resources. Scalability is important to consider, so that simple initial designs can be further built upon to increase throughput or complexity. In initial setups, it is best to start working with a breadboard (a board to set up a temporary circuit that can be easily rearranged) to get all electronics working properly, after which smaller, more solid versions can be created by soldering your circuits. A stepwise approach is advisable for scaling up from simple applications to complex uses. Expansion of existing systems may require some trial-and-error learning and occasionally some replacing of electronic components, but help can easily be sought in one of the many online forums where users provide feedback to all types of questions (Table 2). More advanced solutions can be be to create networks of sensor devices, closed-loop devices, the integration of automatic notifications, live data sharing and visualization, and custom GUIs to control electronics and devices. Open electronics make it also relatively easy to copy and create whole arrays of devices, such as high-throughput recording arrays, weather stations, camera traps, and laboratory monitoring systems (Geissmann *et al.* 2017; Singh *et al.* 2019; Jolles 2020; Tausen *et al.* 2020).

Finally, it is critical to freely share designs, methodologies, and knowledge with the broader community, including a detailed bill of components, fabrication instructions, and photos or illustrations. Beginners can start to publish projects on platforms such as GitHub using Markdown files or on Wikis, which have the benefit of receiving direct feedback from other users, and can then advance to full websites using e.g. Github pages or WordPress. Depending on its novelty one may also decide to write up a methods paper about the specific device and its applications, such as in dedicated open hardware journals (e.g. Journal of Open Hardware or HardwareX).

## 6. Outlook

Potential applications of open electronics are endless and can benefit individual scientists, institutions, and the scientific community as a whole in a broad variety of ways. With the ever-increasing capabilities of electronic components and sensors, and computers becoming more powerful at decreasing size and costs, open electronics are likely to become increasingly used and integrated in our day-to-day life, and over time become a standard component of the scientific toolbox. This in turn will result in new and cutting-edge technologies to be implemented quicker and at a much broader scale, in the lab and in the field. It will also help tech-innovation to expand to other disciplines outside of engineering and thereby fuel the interdisciplinarity of science.

To increase the uptake of open electronics, essential steps are to improve the support by funding organizations, such as to grant researchers extra time in their projects to develop, build and publish open electronics applications and request open hardware alternatives in compulsory instrument bids. Institutions can foster local "ScienceMaker" communities, by providing institutional MakerHubs or workshops, where researchers can prototype and exchange knowledge and ideas with others. Adding electronics and programming training to the institutional career development portfolio would provide further support. Scientific communities can start or join Open Hardware initiatives e.g. Global Open Science Hardware community (Murillo *et al.* 2018), organize dedicated conferences, sessions or workshops to form networks, create standards and foster open electronics across disciplines (Bonvoisin *et al.* 2020).

In this paper we presented the multi-facetted benefits open electronics can offer to researchers, institutions, and the scientific community, to highlight their utility and potential in science. We noted important barriers, and avenues to overcome those - including a beginner's guide. With this we aim to foster a broad uptake of open electronics to support science at multiple scales, from innovation, reproducibility to the democratization of science.

## Acknowledgements

MO was supported by a German Research Foundation fellowship OE 658/2. RLT was supported by a California Sea Grant Delta Science Postdoctoral Fellowship SWC 20-42 Project R/SF-107. TW was supported by the EMBL, the Fellowship Freies Wissen by Wikimedia DE and co, and by ANID/FONDECYT/INICIACION/N°11200666. RS was funded by FEDER through POR Norte (NORTE-01-0145-FEDER-031053) and by National Funds through FCT (PTDC/BIA-BMA/31053/2017 and CEECIND/01424/2017). DO was supported by INTA through projects 2019-PE-E6-I128-001, 2019-PE-E6-I114-001 and REC I117.

## Author contributions

Author contributions modified after the CRediT system (Brand *et al.* 2015).

| Contribution | MO | JWJ | DO | RS | TW | HW | RLT |
|---|---|---|---|---|---|---|---|
| Conceptualization | ▓ | ▓ | ▓ | ▓ | ▓ | | ▓ |
| Investigation | ▓ | | | | | | ▓ |
| Data curation | ▓ | | | | | | |
| Formal analysis | ▓ | | | | | | |
| Writing - outline | ▓ | | | ▓ | ▓ | | ▓ |
| Writing - original draft | ▓ | | | ▓ | ▓ | | ▓ |
| Writing - review and editing | ▓ | ▓ | ▓ | ▓ | ▓ | ▓ | ▓ |
| Visualization | ▓ | | | | | | ▓ |
| Project administration | ▓ | | | | | | ▓ |

## Additional information

Supplementary Information File S1: R Markdown and raw data files for the Web of Science search performed on the 27th April 2021 for the terms "Arduino" and "Raspberry Pi", which included 1866 scientific articles and 7443 conference proceedings. Search fields included title, abstract and keywords. Subject areas were identified by the first ranked subjects identified by the Web of Science for each article. Bibliometric analysis was performed in R (Team 2021) using the bibliometrix package (Aria & Cuccurullo 2017). Files available in the Figshare repository, doi 10.6084/m9.figshare.14875935.

Supplementary Information File S2: R Markdown and raw data files for the PLOS full text analysis of 256 articles, returned by the two search terms "Arduino" and "Raspberry Pi" on 26th May 2021. Scientific disciplines were extracted using PLOS categories and the pubchunks R package (Chamberlain 2020). All analysis was performed in R (Team 2021). Files available in the Figshare repository, doi 10.6084/m9.figshare.14875935.